# Flux period scaling in the Laughlin quasiparticle interferometer

Wei Zhou, F. E. Camino and V. J. Goldman

*Department of Physics, Stony Brook University, Stony Brook, New York 11794-3800, USA*

We report experiments on electron interferometer devices in the quantum Hall regime, where edge channels circle a two-dimensional (2D) electron island. The main confinement is produced by etch trenches, into which front gate metal is deposited. We find a linear dependence of the Aharonov-Bohm period on gate voltage for electrons (integer filling $f = 1$) and for Laughlin quasiparticles (fractional 2/5 embedded in 1/3). The capacitance of a large 2D electron island with respect to the front gates is approximately proportional to the island radius. Comparing the experimental data for the integer and the fractional fillings and for two samples, we find the magnetic field period and its slope scale with the radius of the Aharonov-Bohm orbit. Analysis of the directly measured integer and fractional slope data allows us to determine the interferometer area in the fractional regime, and thus the Laughlin quasiparticle flux period of $5h/e$, within the experimental accuracy.

## I. INTRODUCTION

Recent experiments on electron interferometer devices in the quantum Hall regime, where electron paths circle a 2D electron island, have reported observation of an Aharonov-Bohm superperiod,[1,2] implying fractional statistics of Laughlin quasiparticles.[3-6] Experimental results clearly show Aharonov-Bohm interference of Laughlin quasiparticles in an edge channel of the filling $f = 1/3$ fractional quantum Hall fluid circling an $f = 2/5$ island.[1,2] The experiment determines the magnetic field $B$ period, $\Delta_B$, while the fundamental periodicity of the Aharonov-Bohm effect is as a function the magnetic flux $\Phi = SB$ through a closed path of area $S$. The area $S_{Out}$ of the *electron* Aharonov-Bohm path [Fig. 1(a), the "outer edge ring"] can be determined from the directly-measured Aharonov-Bohm field period $\Delta_B$, using the well-established flux quantization condition $\Delta_\Phi = \Delta_B S = h/e$ for the flux period $\Delta_\Phi$.[7,8] The Aharonov-Bohm area $S_{In}$ enclosed by the inner $f = 1/3$ edge channel (Fig. 1(b), the "$f = 2/5$ island area") is necessarily smaller than $S_{Out}$ in the same device. Thus, unless the flux quantization condition is known *a priori*, the inner area $S_{In}$ can not be deduced from the experimental field period.

The area $S_{In}$ can be evaluated[1] from a self-consistent classical electrostatics mesa edge depletion model of the island electron density profile.[9,8] In addition, it can be evaluated from the tunneling rate estimate, exponentially sensitive to the inner-outer edge separation.[2] However, it could be argued that the microscopic structure of the interferometer edge channels is not known definitively with great accuracy, and thus the modeling of the island electron density profile may possibly involve considerable quantum corrections to the classical electrostatics density profile. Therefore, the relation between the flux $\Delta_\Phi$ and the field $\Delta_B$ periods in the fractional regime is not known accurately from an independent experiment if the $f = 2/5$ island area is not known accurately, which may cast doubt regarding the value of the fractional flux period. Unambiguous determination of $\Delta_\Phi$ is important both for fundamental quantum theory[10] as well as for proposed application of anyons to topological quantum computation.[11-16]

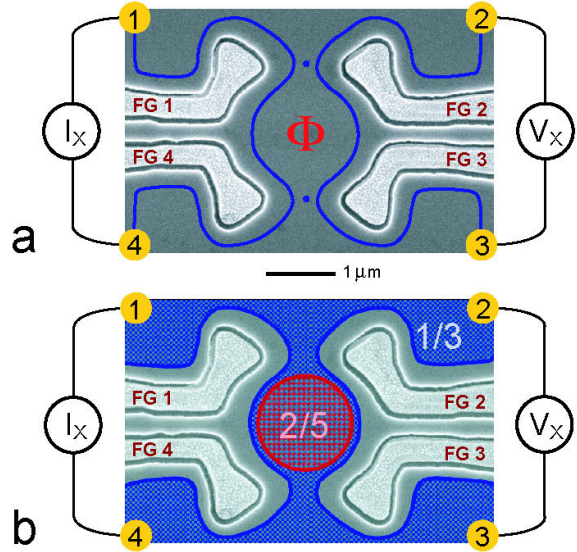

FIG. 1. The Laughlin quasiparticle interferometer samples. (a) and (b) are scanning electron micrographs of a typical device. Four Au/Ti front gates (FG) deposited in shallow etch trenches define the central island separated from the 2D "bulk" by two wide constrictions. The 2D electrons are completely depleted under and near the etched trenches. Four Ohmic contacts are shown schematically by the numbered circles, $R_{XX} \equiv V_{2-3} / I_{1-4}$. The back-gate (not shown) extends over the entire sample on the opposite side of the insulating GaAs substrate. In (a), illustrating quantum Hall filling $f = 1$, the chiral edge channels (blue) follow equipotentials at the periphery of the undepleted 2D electrons; tunneling is shown by dots. (b) illustrates the $f = 2/5$ island surrounded by $f = 1/3$ fractional quantum Hall fluid situation.

Here we report experiments on the dependence of the Aharonov-Bohm field period $\Delta_B$ on the front-gate voltage $V_{FG}$ for electrons ($f = 1$) and for Laughlin quasiparticles (2/5 embedded in 1/3). For moderate $V_{FG}$ we find an approximately linear dependence $d\Delta_B / dV_{FG} = const$ on each quantum Hall plateau. The directly-measured $\Delta_B$ and the slope, $d\Delta_B / dV_{FG}$, and the assumed $S$ can be combined to give $V_{FG}(1e)$, the front-gate voltage attracting charge $1e$ to the area of the A-B orbit.[17] Capacitance is defined as $C \equiv Q/V = e/V_{FG}(1e)$. For a 2D disc of radius $r$, the classical capacitance is approximately proportional to $r$, neglecting a slowly varying logarithmic term. For a large (~2000 electrons) 2D island, the quantum corrections to the classical capacitance are small, and the product $rV_{FG}(1e)$ should be approximately constant, independent of the quantum Hall filling or the area. The island capacitance is also not sensitive to the precise details of edge channel structure since it is an integrated property of the whole island, just like the enclosed A-B flux. Equating the product $rV_{FG}(1e)$ for different quantum Hall regimes, the $f = 2/5$ island area $S = \pi r^2$ can be determined directly with a 10% accuracy. This is quite sufficient to distinguish the physically realistic possibilities of the flux periods $\Delta_\Phi = 5h/e$ (Ref. 1,2), $5h/2e$ (Ref. 18), $h/e$ and $h/2e$ (Ref. 19).

## II. EXPERIMENTAL RESULTS

The quantum electron interferometer samples were fabricated from a low disorder GaAs/AlGaAs heterojunction material where 2D electrons (285 nm below the surface) are prepared by exposure to red light at 4.2 K. The four independently contacted front gates were defined by electron beam lithography on a pre-etched mesa with Ohmic contacts. After a shallow 140 nm wet chemical etching, Au/Ti gate metal was deposited into the etch trenches (lithographic radius $R \approx 1,050$ nm, Sample M97Bm), followed by lift-off, see Fig. 1. Samples were mounted on sapphire substrates with In metal, which serves as a global backgate. Samples were cooled to 10.2 mK; four-terminal resistance $R_{XX} \equiv V_X/I_X$ was measured by passing 50 - 200 pA, 5.4 Hz ac current through contacts 1 and 4, and detecting the voltage between contacts 2 and 3 by a lock-in-phase technique. The four front gates are deposited into etch trenches. Even when front gate $V_{FG} = 0$, the GaAs surface depletion of the etch trenches creates electron confining potential, defining two wide constrictions, which separate an approximately circular 2D electron island from the 2D "bulk", Fig. 2(a).

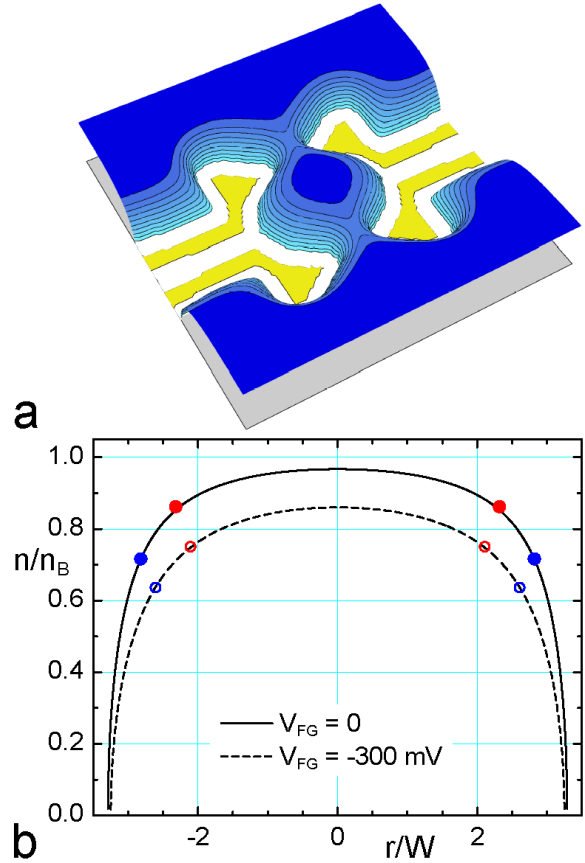

FIG. 2. (a) A qualitative illustration of the 2D electron density profile for the interferometer geometry. (b) The calculated electron density radial profile in a circular island defined by an etched annulus of inner radius $R \approx 1,050$ nm and 2D bulk density $n_B = 1.2 \times 10^{11}$ cm$^{-2}$. The calculation follows the $B = 0$ model of Ref. 8. $W = 245$ nm is the depletion length parameter. The blue circles give the radius of the outer edge ring $r_{Out} \approx 685$ nm, obtained from the integer Aharonov-Bohm period and $n(r_{Out})$ from the $B$-field position of the constriction quantum Hall plateaus. The red circles give the inner edge ring radius $r_{In} \approx 570$ nm, obtained with the fractional flux period $\Delta_\Phi = 5h/e$ and the electron density ratio $n(r_{In})/n(r_{Out}) = (2/5)/(1/3) = 1.20$.

The electron density profile $n(r)$ in a circular island resulting from the etch trench depletion can be evaluated using the model of Ref. 8, based on Ref. 9, see Fig. 2(b). For the bulk density $n_B = 1.2 \times 10^{11}$ cm$^{-2}$, there are ~2,000 electrons in the island. Comparison with a Hartree-Fock profile shows that quantum corrections are significant only for $n < 0.4 n_B$ low density tails,[9] outside the A-B path area. However, the overall density profile follows the $B = 0$ profile in order to minimize the large Coulomb charging energy arising from deviations from the donor-neutralizing $B = 0$ profile.[20] The depletion potential has a saddle point in the constriction region, and so has the resulting electron density profile. From the magnetotransport, we estimate the saddle point density to be $0.72 n_B$. Note that the island center density is slightly (several percent) lower than the 2D bulk density.

On the integer ($f = 1$) and fractional (2/5 embedded in 1/3) quantum Hall plateaus, we acquire the Aharonov-Bohm oscillation data as reported previously.[1,2,8] By varying the front-gate voltage $V_{FG}$, we observe the $B$-field position of the oscillations shift and their period $\Delta_B$ change, see Fig. 3. The effect of the front-gate bias is two-fold. The larger effect is the transistor action affecting the overall 2D electron density in the several micrometer neighborhood of the gates, including the entire island. This is so because at every point in the 2D plane, the electric potential has contributions from the entire (equipotential) front-gate metal area, including the long gate voltage leads, because of the poor screening of the gate electric field by 2D electrons.[20] The overall decrease in the electron density (negative $V_{FG}$) is evidenced by the systematic shift to a lower $B$ of the constriction quantum Hall plateau (with Aharonov-Bohm oscillations superimposed).

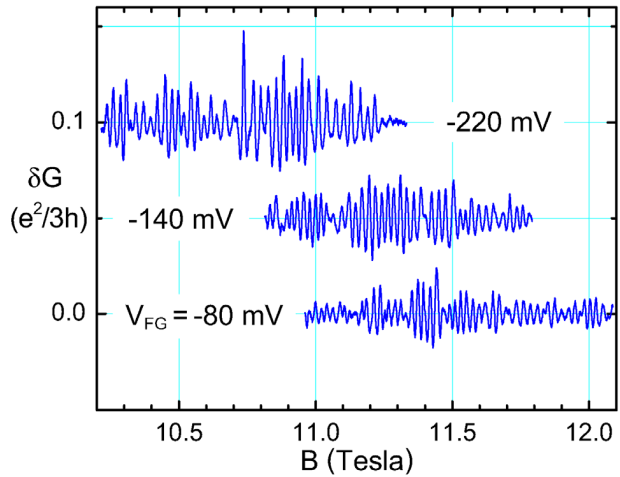

FIG. 3. Aharonov-Bohm conductance oscillations $\delta G$ as a function of $B$ for several values of front gate voltage $V_{FG}$, given in the labels next to each trace. All the traces are for $f = 1/3$ fractional quantum Hall fluid circling an $f = 2/5$ island, and have been shifted vertically in steps of $0.05 e^2 / 3h$. Each trace contains ~40 oscillations with a well-defined period $\Delta_B$, which depends on $V_{FG}$.

In addition, the front gates modify the island and the constriction electron density profile by affecting the primary confining potential of the etch trenches.[8] Since tunneling amplitude is exponentially sensitive to the tunneling distance, the position of the tunneling links at the saddle points in the constrictions is nearly fixed. The constrictions' saddle point electron density determines the equipotential contour of the Aharonov-Bohm path in the island. As evidenced by the systematic increase of the period $\Delta_B$ (decrease of island area, negative $V_{FG}$), the saddle point electron density decreases proportionately less than the island density. Accordingly, remaining on the same quantum Hall plateau, the island edge channels must follow the constant electron density contours with density equal that in the constrictions and move inward, towards the island center, and the A-B path area shrinks. Thus, the electronic charge within the A-B path area decreases because the overall island density decreases, and also because the area itself decreases.

### III. ANALYSIS AND DISCUSSION

The dependence of the Aharonov-Bohm field period $\Delta_B$ on the front gate voltage $V_{FG}$ for electrons ($f = 1$) and for Laughlin quasiparticles (2/5 embedded in 1/3) is shown in Fig. 4. The integer data contains two sets of points from two distinct cooldowns, with ≈7% different 2D electron density, appropriately scaled to produce equal $\Delta_B(V_{FG} = 0)$. We observe an approximately linear dependence $\Delta_B = \Delta_B(0) + (d\Delta_B/dV_{FG})V_{FG}$ in the range of moderate $V_{FG}$ studied; the solid lines are the least squares fits to $\Delta_B = a + bV_{FG}$.

We analyze these data as follows (the analysis aims to express quantities of interest in terms of directly measured quantities and fundamental constants only). As is well known, the Aharonov-Bohm effect is a topological, nonlocal periodic dependence of the phase of a particle's wave function on magnetic flux enclosed by the particle's closed path. In experiment, the interferometer devices are located in a region of a uniform magnetic field, thus observation of A-B effect implies existence of a well-defined closed path that determines the enclosed flux. On the same quantum Hall plateau, the Aharonov-Bohm magnetic flux period is $N_\Phi$ fundamental flux quanta, $\Delta_\Phi = N_\Phi h/e$ (note that the number $N_\Phi$ is not assumed to be an integer here). The Aharonov-Bohm path encloses area $S$ defining the fundamental magnetic flux period

$$\Delta_\Phi = \Delta_B S = N_\Phi h/e, \tag{1}$$

thus $S = N_\Phi h/e\Delta_B$. Differentiating Eq. 1 with respect to front-gate bias, we obtain

$$[(d\Delta_B/dV_{FG})S + (dS/dV_{FG})\Delta_B]_f = 0. \tag{2}$$

The subscript here denotes the same quantum Hall plateau. Substituting $S = N_\Phi h/e\Delta_B$ into Eq. 2 gives

$$dS/dV_{FG} = -(d\Delta_B/dV_{FG})(N_\Phi h/e\Delta_B^2). \tag{3}$$

On the other hand, as is well known, an electron occupies the area $S_1 = 2\pi \ell_0^2 = h/eB$ per spin-polarized Landau level.[21] Thus, Landau level density of electron states is $S_1 = h/eB_1$, where $B_1$ is the magnetic field where the exact filling $\nu = f = 1$ occurs. This expression for $S_1$ can also be obtained by noticing that physically the one-electron area is the inverse of the electron areal density: $S_1 = 1/n$. Recalling that Landau level filling factor $\nu = hn/eB$, we again obtain the one-electron area $S_1 = h/\nu eB = h/eB_1$. We define $V_{FG}(1e)$ as the front-gate voltage required to attract charge $1e$ to the Aharonov-Bohm path area.[17] Linearizing Eq. 3 for $V_{FG}(1e)$, $dS/dV_{FG} = S_1/V_{FG}(1e)$, and substituting $S_1 = h/eB_1$, we obtain

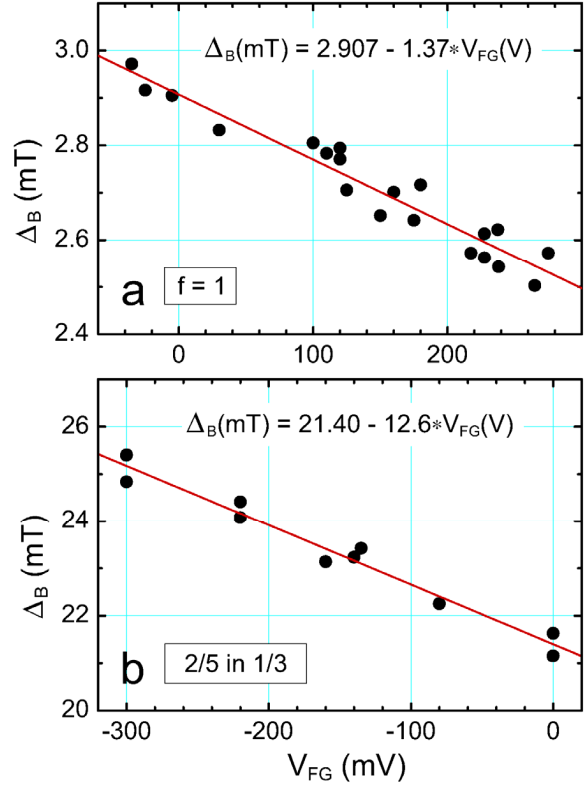

FIG. 4. Dependence of the Aharonov-Bohm period $\Delta_B$ on front-gate voltage $V_{FG}$. The dependence is approximately linear in the range of $V_{FG}$ studied; the solid lines are the least squares fits. The $\Delta_B(V_{FG}=0)$ values give the A-B path areas $S_{Out} = h/e\Delta_B = 1.42 \times 10^{-12}$ m² ($f$ = 1) and $S_{In} = 5h/e\Delta_B = 0.966 \times 10^{-12}$ m² (2/5 embedded in 1/3).

$$V_{FG}(1e) = -\frac{\Delta_B^2}{(d\Delta_B/dV_{FG})N_\Phi B_1}. \qquad (4)$$

Note that the derivative $d\Delta_B/dV_{FG}$ is negative, and that linearization of Eq. 3 is justified by the large size of the electron island, because $1/N_e \approx 1/2000$ is small.

Some discussion on how Eq. 4 applies to the present experimental situation is in order. In the experiment, the sample is located in a uniform magnetic field $B$, which is being slowly varied. The observation of an A-B oscillatory signal thus implies existence of a well-defined A-B area, since the A-B effect is nonlocal and topological in nature and is oscillatory not in $B$, but in magnetic flux through a closed path, which defines the enclosed area (Stokes' theorem). The directly measured period $\Delta_B$ and slope $d\Delta_B/dV_{FG}$ in Eq. 4 refer to the same A-B flux period which is being determined here, so that any visualization of the experimental situation in terms of edge channels is only illustrative and is presented as a physically viable model. The derivation of Eqs. 1-4 does not depend on details of a particular edge channel model used for physical visualization. Thus, Eq. 4 is not sensitive to electron density distribution inside or outside the Aharonov-Bohm path, $N_\Phi$ appears in Eq. 4 only because we express $\Delta_\Phi$ in units of $h/e$.

In an electron system where density is not constant, relation $S_1 = 1/n = h/eB_1$ is still locally valid on a scale of area containing several electrons, that is, several $\ell_0^2$. Density $n$ and the $\nu = 1$ field $B_1$ in Eq. 4 refer to the actual Aharonov-Bohm path, that is, to the edge channel giving rise to the oscillatory conductance signal. In the integer $f = 1$ regime, $n$ and $B_1$ thus refer to the only existing edge channel; $B_1$ (extrapolated to $V_{FG} = 0$) is determined as the field at which $\nu = 1$ occurs for the constriction (not the bulk) QH plateau, on which the $f = 1$ A-B signal is superimposed.[1,8] Thus determined $B_1$ is also used in the fractional (2/5 embedded in 1/3) regime, where it therefore refers to the $f = 1/3$ edge channel, which carries the transport current (Hall resistance plateau[22] is at $3h/e^2$), and where the fractional A-B signal originates. This is justified because the edge channel giving rise to the A-B signal must pass through the constrictions, and the corresponding density is thus determined by the saddle point density in the constrictions, as discussed in Sec. II. Thus, the physical interpretation of Eq. 4 is that it describes effective electrostatic coupling to the front gates of electrons located at the position of the A-B path.

The experimentally determined field periods $\Delta_B(V_{FG} = 0)$, their slopes $d\Delta_B/dV_{FG}$, and $B_1$ for two samples are summarized in Table I. Equation 4 is used to obtain $V_{FG}(1e)$ and the product $rV_{FG}(1e)$ assuming $N_\Phi = 5$ for the fractional quantum Hall regime, for the Aharonov-Bohm path within the $f = 1/3$ edge ring enclosing the 2/5 island. The integer $N_\Phi = 1$; the A-B path radius $r$ assumes a circular path, alternatively $\sqrt{S}$ can be used instead of $r$. Note that using $N_\Phi = 5$ for the inner 2/5 in 1/3 island gives roughly equal $V_{FG}(1e)$; the products $rV_{FG}(1e)$ are equal (within the experimental uncertainty of $\pm 10\%$), as expected. Assuming different $N_\Phi$ gives correspondingly different fractional $V_{FG}(1e) \propto 1/N_\Phi$ and $rV_{FG}(1e) \propto 1/\sqrt{N_\Phi}$, inconsistent with the expectation. For example, using the next physically feasible flux period $\Delta_\Phi = 5h/2e$ ($N_\Phi = 2.5$) increases the fractional regime value of $V_{FG}(1e)$ by 2, and the value of $rV_{FG}(1e)$ by $\sqrt{2}$, well outside of the experimental uncertainty. Using $\Delta_\Phi = h/2e$, corresponding to excitation of one $e/5$ quasiparticle in the 2/5 island, yields $rV_{FG}(1e) = 3.26$ V·nm, implying 3.2 times weaker coupling of the island electrons to the front gates, whereas an approximately constant coupling is expected from the gate geometry. We thus conclude that values of $N_\Phi \leq 2.5$ are not consistent with the experimentally observed $\Delta_B$ and $d\Delta_B/dV_{FG}$.

Alternatively, without explicitly using $N_\Phi$, we can rewrite the Eqs. 1-4 in terms of $S_{Out} = h/e\Delta_B$ from $f = 1$ and $S_{In}$ via the directly measured $\Delta_B$ and $d\Delta_B/dV_{FG}$. Requiring exact equality of the products $V_{FG}(1e)\sqrt{S_{In}}$ and $V_{FG}(1e)\sqrt{S_{Out}}$ obtained from the fractional and the integer data, respectively, we obtain an equation for $S_{In}$:

$$\sqrt{\frac{h}{e}} \frac{1}{B_1} \left[ \frac{\sqrt{\Delta_B^3}}{d\Delta_B/dV_{FG}} \right]_{f=1} = \frac{h}{eB_1} \left[ \frac{\Delta_B}{\sqrt{S_{In}}(d\Delta_B/dV_{FG})} \right]_{f=2/5}. \tag{5}$$

This gives

TABLE I. Summary of results obtained from the experimental Aharonov-Bohm period $\Delta_B$ and its dependence on front gate voltage $V_{FG}$ as described in the text. Sample M61Dd data is from Ref. 8.

| Sample | M97Bm $f=1$ | M97Bm 2/5 in 1/3 | M61Dd $f=1$ |
|---|---|---|---|
| $\Delta_B(0)$, mT | 2.907 | 21.40 | 1.872 |
| $d\Delta_B/dV_{FG}$, mT/V | $-1.37$ | $-12.6$ | $-1.22$ |
| $B_1$, T | 3.92 | 3.92 | 2.53 |
| $V_{FG}(1e)$, mV | 1.58 | 1.86 | 1.14 |
| $r$, nm | 673 | 555 | 839 |
| $rV_{FG}(1e)$, V·nm | 1.06 | 1.03 | 0.956 |

$$S_{In} = \frac{h}{e}\left[\frac{d\Delta_B/dV_{FG}}{\sqrt{\Delta_B^3}}\right]^2_{f=1} \times \left[\frac{\Delta_B}{d\Delta_B/dV_{FG}}\right]^2_{f=2/5}, \tag{6}$$

yielding $S_{In} = 0.92 \times 10^{-12}$ m$^2$ and $\Delta_\Phi = \Delta_B S_{In} = 2.0 \times 10^{-14}$ Wb $= 4.8 h/e$ from the data of Fig. 4, with an experimental uncertainty of ±10%. The dominant source of experimental error is the uncertainty in the $\Delta_B$ vs. $V_{FG}$ slopes.

Although we do not use the electron density modeling in the data analysis presented above, it is interesting to compare the qualitative features of the front-gate bias-dependence of the oscillatory data of the kind presented in Fig. 3 to the calculated island electron density profile, Fig. 2(b). The open and closed circles show the $n(r)$ positions obtained from the integer and the fractional Aharonov-Bohm data for $V_{FG} = 0$ and $-300$ mV, respectively. The fractional regime circles (red) give radii using the flux period $\Delta_\Phi = 5h/e$. Only one ($f=1$, $V_{FG}=0$, blue solid circle) of the four independent points is adjusted to fit the experiment, thus "calibrating" the depletion length parameter $W$. Both effects observed experimentally are consistent with the profile of Fig. 2(b): the systematic shift of the same filling factor upon application of $V_{FG}$, and the systematic change of the Aharonov-Bohm oscillation period. It is also worth mentioning that a stable edge ring requires steep enough gradient of the confining potential, that is, steep $-e(\partial n/\partial r)$. The experimental fact that the fractional quantum Hall regime Aharonov-Bohm oscillations persist even upon application of a moderate $V_{FG} = -300$ mV rules out inner edge ring radii well inside the island, where the confining potential gradient is very small at $V_{FG}=0$, so that application of a moderate negative $V_{FG}$ would shrink the A-B orbit to zero.

## IV. CONCLUSIONS

In conclusion, we report experiments on electron interferometer devices in the quantum Hall regime, focusing on determination of the Aharonov-Bohm magnetic field period $\Delta_B(V_{FG}=0)$ and its front-gate voltage slope $d\Delta_B/dV_{FG}$ for electrons (quantum Hall filling $f=1$) and for Laughlin quasiparticles (2/5 embedded in 1/3). The Aharonov-Bohm period and its derivative can be combined to give the increment of the gate voltage attracting charge $1e$, that is, the electrostatic coupling of electrons to the front gates, assuming the area of Aharonov-Bohm orbit is known. This allows us to determine the fractional quantum Hall regime flux period $\Delta_\Phi$ directly, without reference either to a calculated electron density profile, or the tunneling distance consistent with the experimental amplitude of conductance oscillations. We find the fractional flux period $\Delta_\Phi = 5h/e$ is consistent, while $\Delta_\Phi \leq 2.5h/e$ are inconsistent with the front-gate bias experimental results.

### ACKNOWLEDGMENTS

We thank D. V. Averin for discussions. This work was supported in part by U.S. NSA and ARDA through U.S. Army Research Office under Grant DAAD19-03-1-0126 and by the National Science Foundation under Grant DMR-0303705.